\documentclass[a4paper,10pt]{article}
\usepackage[bf]{caption} 

\usepackage{graphicx}
\usepackage{hyperref}

\usepackage{txfonts} %Please comment out this line unless the txfonts package is availabe in your LaTeX system.

\begin{document}

%\vspace*{0.cm}
\title{\vspace*{-0.7cm}
\raggedright
\bf \LARGE Plastic scintillator detector with the readout based on an array of large-area SiPMs for the ND280/T2K upgrade and SHiP experiments
%\vspace{0.1cm}
}

\author{
\normalsize\hspace{-2mm}A.\,Korzenev$^{1}$, C.\,Betancourt$^{2}$, A.\,Blondel$^{1}$, D.\,Breton$^3$, A.\,Datwyler$^{2}$, D.\,Gascon$^{4}$, \hspace{\fill} \\
\normalsize \hspace{-2mm}S.\,Gomez$^{4}$, M.\,Khabibullin$^{5}$, Y.\,Kudenko$^{5,6,7}$, J.\,Maalmi$^3$, P.\,Mermod$^{1}$, E.\,Noah$^{1}$, N.\,Serra$^{2}$, \hspace{5cm} \\ 
\normalsize\hspace{-2mm}D.\,Sgalaberna$^{8}$,  B.\,Storaci$^{2}$ \hspace{\fill} 
}

\date{}
\maketitle

\vspace{-0.7cm}
\noindent
{\it
$^{1}$ DPNC, University of Geneva, Geneva, Switzerland\\
$^{2}$ Physik-Institut, Universit\"{a}t Z\"{u}rich, Z\"{u}rich, Switzerland\\
$^{3}$ Laboratoire de L'acc\'{e}l\'{e}rateur Lin\'{e}aire from CNRS/IN2P3, Orsay, France\\
$^{4}$ Institut de Ci\`{e}ncies del Cosmos, Universitat de Barcelona, Barcelona, Spain\\
$^{5}$ Institute for Nuclear Research of the Russian Academy of Sciences, Moscow, Russia\\
$^{6}$ Moscow Institute of Physics and Technology, Moscow Region, Russia\\
$^{7}$ Moscow Engineering Physics Institute (MEPhI), Moscow, Russia\\
$^{8}$ European Organization for Nuclear Research (CERN), Geneva, Switzerland
}
\vspace{0.2cm}
%\thanks{alexander.korzenev@cern.ch}

\noindent
{\it E-mail: alexander.korzenev@cern.ch}
\vspace{0.7cm}
%\recdate{October 5, 2016}

\noindent
%\begin{abstract}
\begin{minipage}{0.9\textwidth}
\small
Plastic scintillator detectors have been extensively used in particle physics experiments 
for decades.
A large-scale detector is typically arranged as an array
of staggered long bars which provide a fast trigger signal and/or 
particle identification via time-of-flight measurements.
Scintillation light is collected by photosensors coupled to both ends of every bar.
In this article, we present our study on a direct replacement of commonly used 
vacuum photomultiplier tubes (PMTs) by arrays of large-area silicon photomultipliers (SiPMs). 
An SiPM array which is directly coupled to the scintillator bulk, 
has a clear advantage with respect to a PMT: compactness, mechanical robustness, 
high PDE, low operation voltage, insensitivity to magnetic field, 
low material budget, possibility to omit light-guides. 
In this study, arrays of eight $6 \times 6$ mm$^2$ area SiPMs were coupled to the ends 
of  plastic scintillator bars with 1.68 m and 2.3 m lengths.
An 8 channel SiPM anode readout ASIC (eMUSIC) was used for the readout, 
amplification and summation of signals of individual SiPMs.
Timing characteristics of a large-scale detector prototype were studied in 
test-beams at the CERN PS.
This technology is proposed for the ToF system of the ND280/T2K\,II upgrade at J-PARC 
and the timing detector of the SHiP experiment at the CERN SPS.
\end{minipage}
%\end{abstract}
\vspace{0.3cm}
%\kword{MPPC, SiPM array, Plastic scintillator, Timing detector, ToF, SHiP, ND280, T2K}

\noindent
\textsf{{\bf\small \textsf{KEYWORDS:}} \small MPPC, SiPM array, Plastic scintillator, Timing detector, ToF, SHiP, ND280, T2K
}
\vspace{0.5cm}
%\maketitle

%%%%%%%%%%%%%%%%%%%%%%%%%%%%%%%%%%%%%%%%%%%%%%%%%%%%%%%%%
%%%%%%%%%%%%% Introduction %%%%%%%%%%%%%%%%%%%%%%%%%%%%%%

\section{\large Introduction}

~~~~~A study of timing properties of a %long plastic scintillator counter 
detector assembled from long plastic scintillator bars which are read out by
silicon photomultiplier
(SiPM) arrays is presented.
SiPMs are widely employed in high energy physics detectors.
The fast evolution of the SiPM market opens new possibilities.
Large-area sensors are nowadays available in the market at a reasonable price.
When assembled in a 2D array they can cover a sizeable area,
therefore they can be considered as a promising direct replacement for traditional 
photomultiplier tubes (PMTs).
In particular, the sensors can be coupled directly to a scintillator bulk
to provide a time resolution on a sub 100~ps level \cite{JINST}.
An obvious advantage of an SiPM array is that it can take a form of
the bar cross-section, thus avoiding complex shape adiabatic light-guides.

A large SiPM capacitance increases the rise time and width of a signal and 
thus worsens the time resolution. 
In this regard, a large monolithic sensor or many smaller sensors with common cathode
and anode cannot be employed.
%A reduction of the overall capacitance
An improvement can be achieved by using an independent sensor readout to isolate the sensor capacitances from each other. 
Signals can be amplified and summed afterwards either by a discrete circuit \cite{Jung}
or by an ASIC without the drawback of adding the sensor capacitances at the input. 
The eMUSIC ASIC \cite{MUSIC} was adopted as an input stage of the acquisition system. % used in this study.

%The test-bench used in this work can be considered as a prototype for the design of 
%the timing detector of the SHiP experiment \cite{SHiP} and  the ToF system proposed for the ND280/T2K upgrade \cite{Davide}. 
%Furthermore, the test-bench is used as a test ground for a recently developed ASIC MUSIC\,R1 \cite{MUSIC} which was employed in a test-beam study for the first time.

A detector configuration where SiPM arrays are coupled to long 
plastic scintillator bars, signals of SiPMs are read out by the eMUSIC chip
and, finally, analog signals are digitized by a Waveform and TDC converter (WTDC) 
is proposed for the time-of-flight (ToF) system of the ND280/T2K\,II upgrade~\cite{Davide} at J-PARC 
and the timing detector of the SHiP experiment \cite{SHiP} at the CERN SPS.

\begin{figure}[t]
\centering
\includegraphics[width=\textwidth]{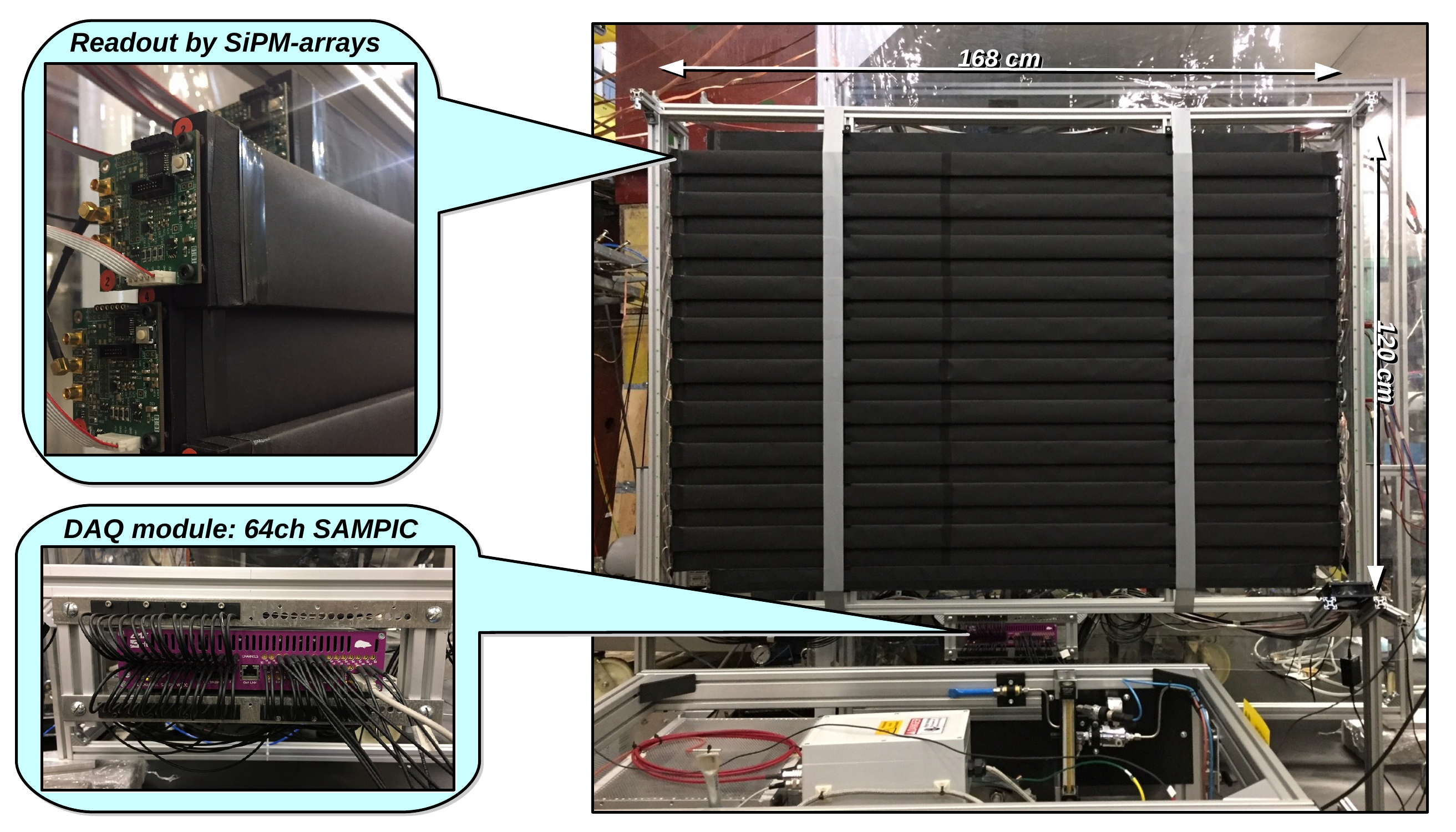}
\caption{A front view of the prototype detector which was assembled in summer 2018.
  Two inserts on the left-hand side represent a zoomed view of
  the eMUSIC readout board ({\it top}) and
  the 64 channel SAMPIC data acquisition module ({\it bottom}).
}
\label{fig:proto_photo}
\end{figure}

%%%%%%%%%%%%%%%%%%%%%%%%%%%%%%%%%%%%%%%%%%%%%%%%%%%%%%%%%%
%%%%%%%%%%%%%% Prototype %%%%%%%%%%%%%%%%%%%%%%%%%%%%%%%%%

\section{\large Large-scale prototype}

~~~~A 22 bar prototype, shown in Fig.\,{\textmd {\ref{fig:proto_photo}}}, 
%very similar to one of the planes of the ND280 TOF detector
was assembled and exposed for one month to test beams of the CERN PS in summer 2018.
%This prototype operated successfully demonstrating the high performance at relatively low cost, as well as the viability of the proposed solutions for the power distribution and heat dissipation, the signal readout, the DAQ using a SAMPIC waveform digitiser, the time synchronisation between bars, and the integration with other sub-detectors. 
In addition to the test of operating performance of the prototype itself
it was also used as a time-of-flight detector for the identification of particles
with momenta up to 6~GeV/$c$. 
The particles were identified via measurements of the time difference between the prototype
and a beam counter, S1, installed 10.9~m upstream.
%The latter was plastic scintillator of $4 \times4$ cm$2$ cross-section installed 10.9~m upstream.

The choice of  scintillator material was driven by the need to achieve precision timing 
by detecting as many photons as possible for 
interactions occurring all along the full length of a bar. 
The scintillator EJ-200 provides an optimal combination of a high light output, 
suitable optical attenuation length of about 4~m, and fast timing 
(rise time of 0.9 ns).

\begin{figure}[t]
\centering
\includegraphics[width=\textwidth]{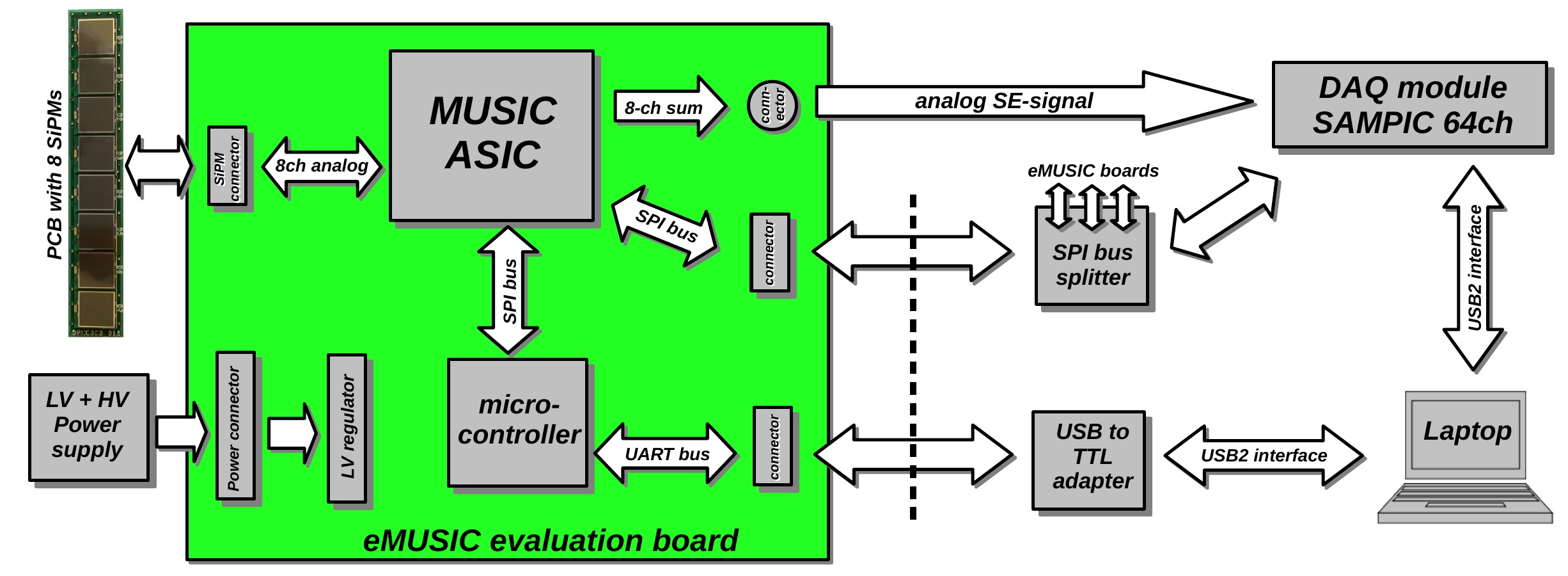}
\caption{A block diagram of the eMUSIC mini-board  profiles
  ({\it left-hand side})
  and a schematic representation of a data flow ({\it right-hand side}).
}
\label{fig:block_diagram}
\end{figure}

An array of 8 surface-mount devices S13360-6050PE 
(area $6\times6$~mm$^2$, pixel pitch 50~$\mu$m) from Hamamatsu has been mounted 
to a custom-made PCB, as shown %on top of plots in Fig.\,\ref{fig:time_resolution}.
in Fig.\,\ref{fig:block_diagram} (left).
A charge produced in the array was sent via a high-density connector to
a general purpose mini-board %eMUSIC v2 
which was based on the eMUSIC chip \cite{MUSIC} providing 
either an individual SiPM readout or an analog sum of all eight SiPM channels.
The outputs of the chip could be easily monitored via coaxial RF connectors
as a differential or single-ended signals.
The eMUSIC chip has several configuration parameters accessible via an SPI protocol,
i.e.~a tunable pole-zero cancellation providing output signals 
with less than 10~ns FWHM; two different gain options; 
a bias voltage of every SiPM could be controlled using an internal
DAC with 1\,V dynamic range; 
any of channels could be powered-off. 

A block diagram of the eMUSIC mini-board is presented in Fig.\,\ref{fig:block_diagram}. 
There are two ways to modify and upload parameters to the ASIC. 
%One can use
%\begin{itemize}
\begin{enumerate}
\item  A limited programmability profile (UART).
  In this case, the eMUSIC board has to be connected to a PC
  via an external UART-to-USB adapter. An in-board microcontroller is
  used as a bridge to send configurations to a non-volatile memory of EEPROM.
  The microcontroller dumps this configuration to the eMUSIC chip 
  as soon as the board has been switched on. 
\item  A full programmability profile (SPI).
  In this mode, the embedded microcontroller is bypassed and 
  the external master takes control of the SPI bus.
  This profile is suitable for applications where a frequent reconfiguration of
  the eMUSIC chip is required. In this mode, the control
  is taken by an FPGA of the DAQ module. 
  An SPI bus from multiple eMUSIC mini-boards is propagated to the DAQ module
  using SPI-splitter boards.
%\end{itemize}
\end{enumerate}

A 64 channel SAMPIC module~\cite{SAMPIC} was used to digitize the signals (see Fig.\,\ref{fig:proto_photo}). 
%It is an acquisition electronics which performs a waveform sampling using a switched capacitor array. % (SCA).
SAMPIC is a 16\,channel ASIC implementing a novel type of digitizing electronics which performs both the function of a TDC and a waveform sampler based on a switched capacitor array.
%A use of an analogue memory which is added in parallel with a delay line allows for analog signal sampling at a very high rate. In addition,
Having the waveforms recorded, one can extract various kinds of information such as
baseline, amplitude, charge and time. 
The SAMPIC circular buffer contains 64 cells which makes possible
to cover a 20 ns time window at the sampling frequency of 3.2 GS/s. 
%The ASIC contains an on-chip ADC thus digitizes both time and charge. 
In addition to the TDC, the ASIC contains one on-chip ADC per cell which makes the charge digitization particularly fast.
Each SAMPIC channel integrates a discriminator which can trigger itself
independently of other channels. It is an important feature for neutrino experiments which do not have triggers induced by  incoming particles.
% (incoming particle is not detected).
%The data flow is schematically presented in Fig.\,\ref{fig:block_diagram}.

\begin{figure}[t]
\centering
\includegraphics[width=\textwidth]{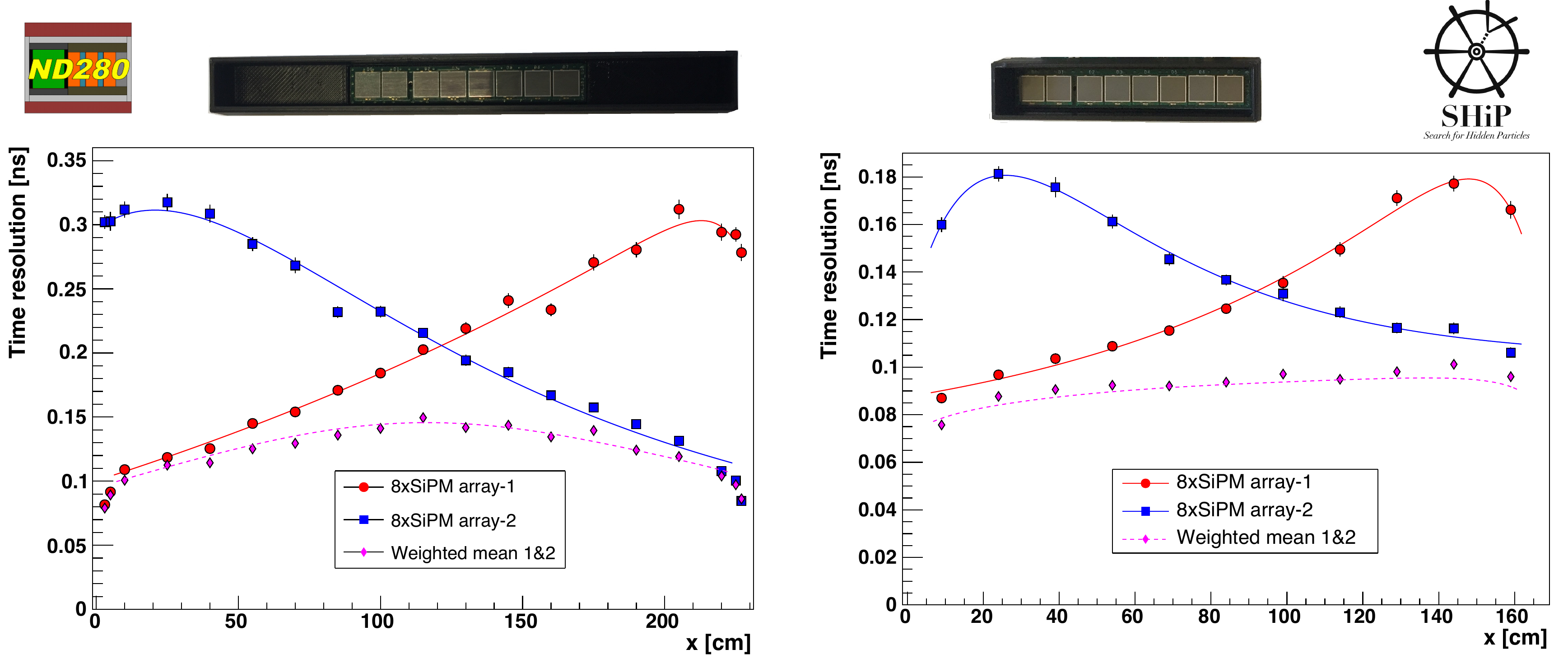}
\caption{Time resolution as measured by 8-SiPM arrays attached at both ends of
  a $230\times12\times1$~cm$^3$ bar of ToF detector of ND280 ({\it left}) and
  a $168\times6\times1$~cm$^3$ bar of timing detector of SHiP ({\it right})
  as a function of the interaction point along the bar.
  Arrays of eight SiPMs  fixed inside of
  light-tight cases are shown on top of corresponding figures.
  The sensitive area of each SiPM is $6 \times 6$ mm$^2$.
}
\label{fig:time_resolution}
\end{figure}

The time resolution of bars with dimensions 
$230\times12\times1$~cm$^3$ and $168\times6\times1$~cm$^3$
%as registered by the arrays at their both ends
is shown in Fig.\,\ref{fig:time_resolution}.
The technique of the measurements is described in Ref.\,\cite{JINST}.
The resolution evolves from 80~ps for the crossing point located near the sensor, 
to 310~ps (230~cm bar) and 180~ps (168~cm bar) for the
% light propagation along the full length of the bar.
crossing point located at the opposite end of the bar to that of the sensor, due to light propagation along the full length of the bar.
The resolution, calculated as a weighted mean between SiPM-arrays located at two ends
of each bar, makes the distribution more constant, i.e.~130~ps and 85~ps on average for two bars, respectively.

\begin{figure}[t]
\centering
\includegraphics[width=0.48\textwidth]{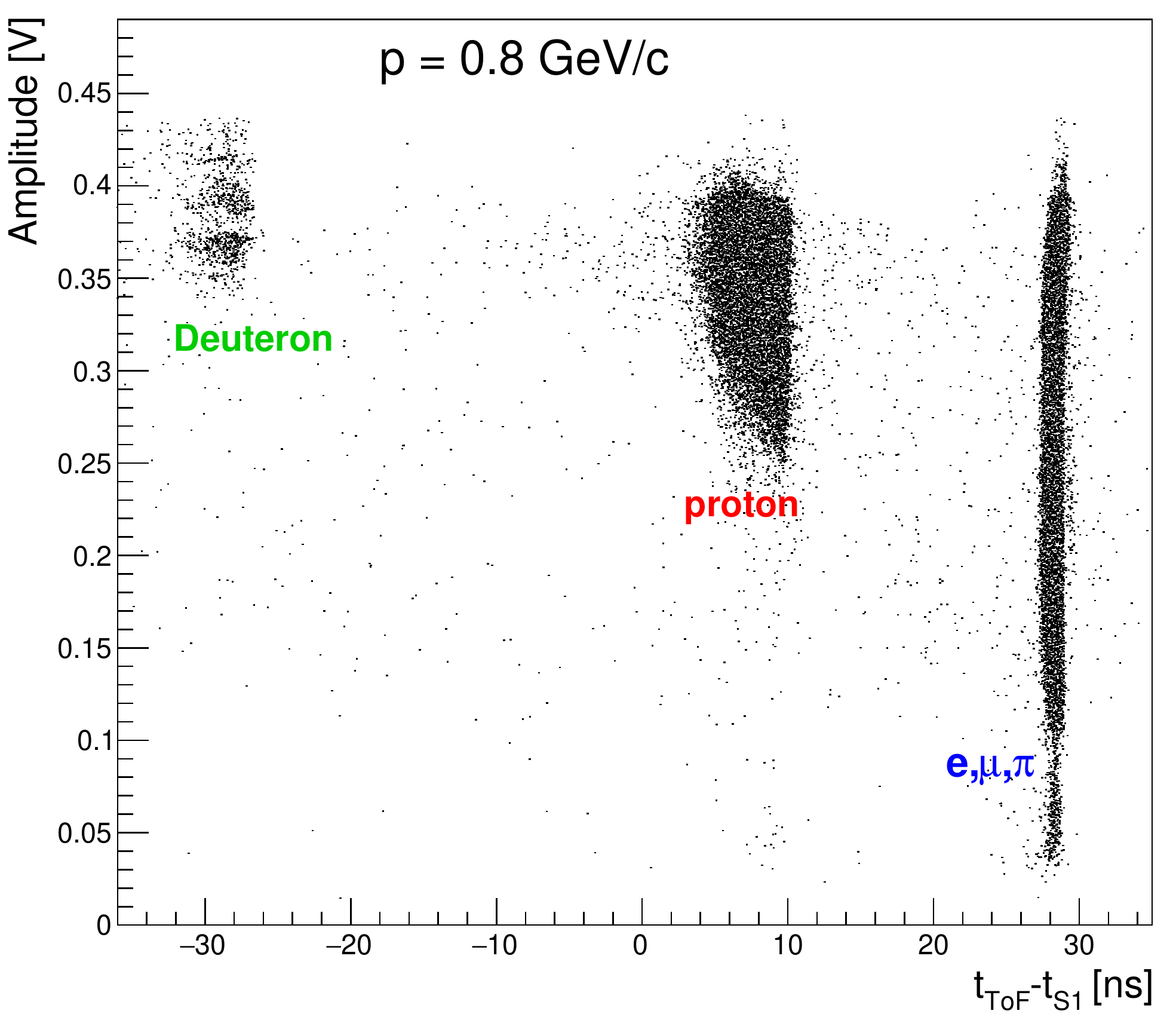}
\hfill
\includegraphics[width=0.47\textwidth]{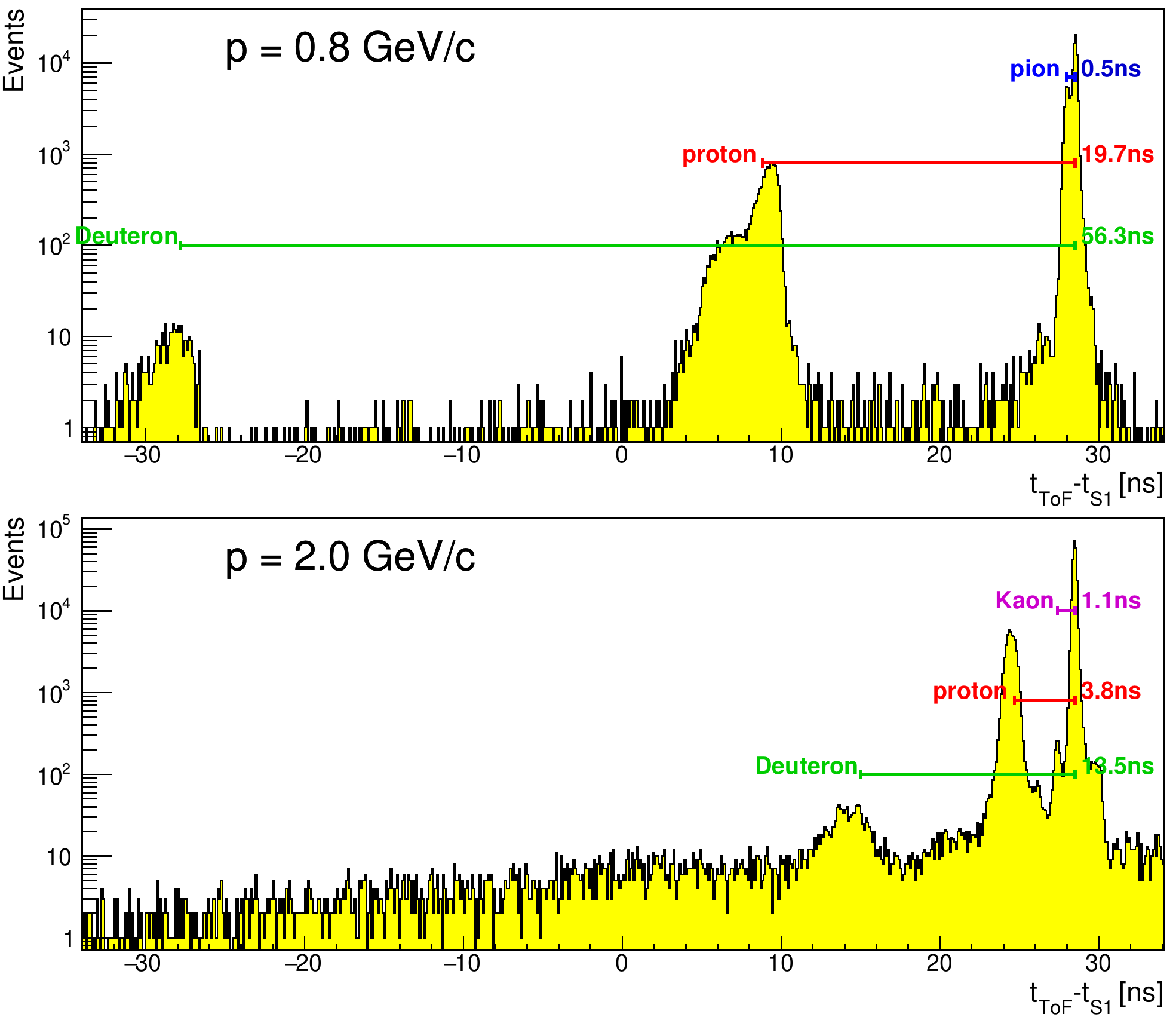}
\caption{{\it Left}: a correlation plot presenting a signal amplitude versus time
  required by 0.8~GeV/$c$ particles to traverse a 10.9~m distance between 
  the beam counter S1 and the ToF prototype.
  {\it Right}: a time required by 0.8~GeV/$c$ and 2~GeV/$c$ particles to traverse 
  the distance between S1 and the ToF prototype.
  }
  \label{fig:hists}
\end{figure}

Some results obtained with the 22 bar prototype are presented in Fig.\,\ref{fig:hists}.
The primary goal was to identify particles by calculating the time
difference detected by the S1 counter and the prototype.
A longer time was required by heavier particles to traverse this distance 
which could serve to identify the species. As an example, protons
arrive at the prototype about 20~ns later as compared to positrons at 0.8~GeV/$c$.
This interval shrinks to 3.8~ns at 2~GeV/$c$.

An ionization produced by charged particles and, in turn, an amount of
optical photons generated in de-excitation processes depends on particle masses 
and is quite different at low momenta. 
The effect can be observed in the amplitude of recorded signals 
as shown in Fig.\,\ref{fig:hists} (left).
Amplitudes of the deuteron and proton signals ranges around 300 -- 400 mV
whereas signals of e, $\mu$ and $\pi$ are spread from 50 to 400 mV.
This information can also be used for the particle identification.

The technology described in this article has been proposed for 
%the time-of-flight system of the upgrade programme of the near detector of T2K \cite{Davide} at J-PARC and for the timing detector of the SHiP experiment \cite{SHiP} at the CERN SPS. 
two time measuring detectors which are presented in the following sections.
Both detectors are located in zones of strong magnetic field.

%%%%%%%%%%%%%%%%%%%%%%%%%%%%%%%%%%%%%%%%%%%%%%%%%%%%%%%%%%
%%%%%%%%%%%%%% ND280/T2K %%%%%%%%%%%%%%%%%%%%%%%%%%%%%%%%%

\section{\large Time-of-flight detector for the ND280/T2K\,II upgrade}

~~~~The goal of the T2K experiment is to study neutrino flavor oscillations employing
an off-axis neutrino beam from the J-PARC accelerator facility.
The near detector of T2K, ND280, is used to study neutrino interactions
aiming for neutrino cross-section measurements and 
reduction of systematic uncertainties in neutrino oscillation analyses.
%Super-Kamiokande is used as a far detector.

A layout of the ND280 detector proposed for the upgrade of T2K \cite{Davide} and 
the ToF detector itself are shown in Fig.\,\ref{fig:fig_ND280}.
The ToF system will consist of 6 planes surrounding the active target 
\cite{Mineev:2018ekk} and two TPCs. 
Each plane will have about 5~m$^2$ surface area. 
The ToF aims at precise measurements of the crossing time of charged particles 
as they exit or enter the TPCs.
A time resolution of 500~ps or better is required for an unambiguous 
determination of the flight direction of charged particles. 
An additional goal is to improve particle identification, 
requiring the time resolution of 100 -- 200~ps. 
%The TOF should be based on a technology which is cheap, compact, efficient, robust and magnetic-field tolerant. The solution described below meets all these requirements with a 150~ps time resolution as shown in Fig.\,\ref{fig:time_resolution}\,(left).
%
The whole system will be located inside the UA1 magnet which will create 
a field of 0.2~T. 
The presence of the magnetic field and a very limited space for the detector
makes the use of the SiPM readout particularly advantageous.
%The main purpose of the detector is to determine the trajectory direction of particles (inside or outside the target) which are products of neutrino interactions.
%Since the energy of the neutrino beam peaks at 0.6 GeV the momentum of secondaries resides in a hundred-MeV region which also makes possible to use the detector for the particle identification although the flight distance between planes is short.
The detector will be assembled from 118 bars with a length of 2 to 2.3~m
which will provide a time resolution of approximately 150~ps along the bar
as shown in Fig.\,\ref{fig:time_resolution} (left).
In total, the number of DAQ channels is 236 and the number of SiPMs is 1888.

%\begin{figure}[t]
%\centering
%\includegraphics[width=\textwidth]{ToF_ND280_all6}
%\caption{Layout of the ND280 detector proposed for the T2K\,II upgrade, with magnets
%  opened such as to see the inner basket ({\it right}).
%  The part of the basket to be upgraded is shown {\it  on the left}.
%  It includes the active target and two TPCs,
%  all surrounded by 6 ToF planes.
% }
%\label{fig:fig_ND280}
%\end{figure}

%%%%%%%%%%%%%%%%%%%%%%%%%%%%%%%%%%%%%%%%%%%%%%%%%%%%%%%
%%%%%%%%%%%%%% SHIP %%%%%%%%%%%%%%%%%%%%%%%%%%%%%%%%%%%%%%

\section{\large Timing detector of SHiP}

~~~~~SHiP is a new general-purpose experiment \cite{SHiP} proposed for installation at a beam dump facility
of the CERN SPS to search for hidden particles as predicted by a very large number of
recently elaborated models of Hidden Sectors which are capable of accommodating dark matter,
neutrino oscillations, and the origin of the full baryon asymmetry in the Universe.

%Specifically, the experiment is aimed at searching for very weakly interacting long lived particles including Heavy Neutral Leptons - right-handed partners of the active neutrinos, vector, scalar, axion portals to the Hidden Sector, and light supersymmetric particles - sgoldstinos, etc.

\begin{figure}[t]
\centering
\includegraphics[width=\textwidth]{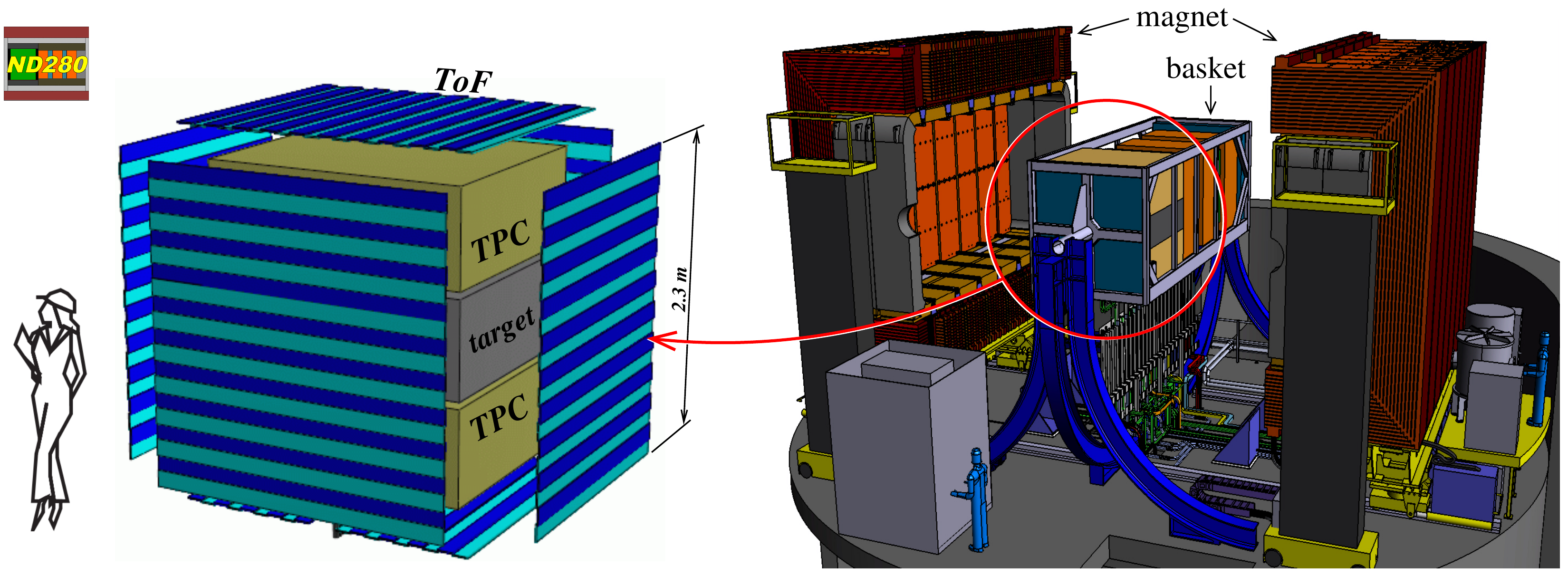}
\caption{Layout of the ND280 detector proposed for the T2K\,II upgrade, with magnets
  opened such as to see the inner basket ({\it right}).
  The part of the basket to be upgraded is shown {\it  on the left}.
  It includes the active target and two TPCs,
  all surrounded by 6 ToF planes.
 }
\label{fig:fig_ND280}
\vspace{0.6cm}
\centering
\includegraphics[width=\textwidth]{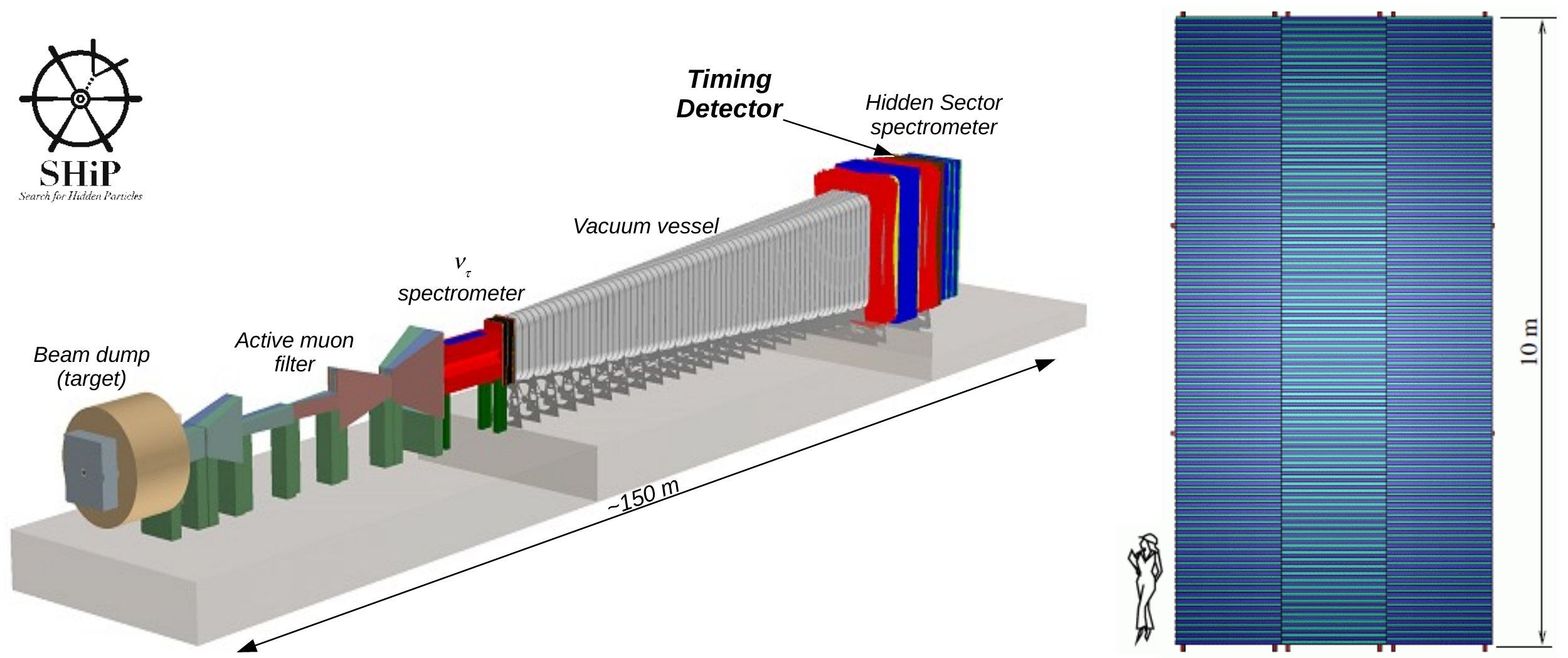}
\caption{{\it Left}: the SHiP experimental facility.
  {\it Right}: schematic view to the SHiP timing detector 
  which will be located downstream of the vacuum vessel,
  in front of an electromagnetic calorimeter.
}
\label{fig:fig_SHIP}
\end{figure}

The SHiP spectrometer and the timing detector are shown in Fig.\,\ref{fig:fig_SHIP}.
The timing detector will be positioned downstream of the vacuum decay vessel and
will  cover a $5 \times 10$ m$^2$ area.
The main purpose of the detector is a reduction of the combinatorial background
(vertices made by a random muon crossing)
by tagging particles belonging to a single event.
The time resolution is required to be better than 100~ps.
The detector can also be used for identification of few~GeV particles.
The 100~ps resolution constraint limits the possible bar length to approximately 2~m.
Therefore an array of 3 columns and 182 rows assembled
from 167 cm long bars is considered
as a base option for the present design of the detector.
Altogether, it results in 546 bars, 1092 channels and 8736 SiPMs.
The test-beam measurements for a single bar resulted in about 85~ps time resolution
as shown \mbox{in Fig.\,\ref{fig:time_resolution} (right).}

%%%%%%%%%%%%%%%%%%%%%%%%%%%%%%%%%%%%%%%%%%%%%

\section{\large Conclusions}

A study on a direct replacement of vacuum photomultiplier tubes 
by arrays of large-area silicon photomultipliers has been presented.
Arrays of SiPMs were coupled to the ends of long plastic scintillator bars. 
 An 8\,channel chip, eMUSIC, was used for the readout, amplification and summation of signals of individual SiPMs.
 A 64\,channel SAMPIC module was used for the data acquisition.
Results obtained in test-beams with a 22 bar prototype detector have been presented.
%Timing characteristics of a large-scale detector prototype were studied in test-beams at the CERN PS. 
This technology is proposed for the ToF system of the ND280/T2K II upgrade at J-PARC and the timing detector of the SHiP experiment at the CERN SPS.

%%%%%%%%% reference %%%%%%%%%%%%%%%%%%%%%%%%%%%%%

{\small
\bibliography{proc_PD18_korzenev} 

\providecommand{\href}[2]{#2}\begingroup\raggedright\begin{thebibliography}{1}

\bibitem{JINST}
C.~Betancourt et~al., \emph{{Application of large area SiPMs for the readout of
  a plastic scintillator based timing detector}},
  \href{https://doi.org/10.1088/1748-0221/12/11/P11023}{\emph{JINST} {\bfseries
  12} (2017) P11023}, [\href{https://arxiv.org/abs/1709.08972}{{\ttfamily
  1709.08972}}].

\bibitem{Jung}
W.~Jung et~al., \emph{{Development of TPC Trigger Hodoscope for J-PARC E42/E45
  hadron experiment}},  in \emph{{{International Workshop on New Photon
  Detectors (PD18): Tokyo, Japan, November 27-29, 2018}}}, 2018.

\bibitem{MUSIC}
S.~Gomez et~al., \emph{{MUSIC: An 8 channel readout ASIC for SiPM arrays}},  in
  \emph{{Proceedings, Optical Sensing and Detection IV, SPIE Photonics Europe,
  2016, Brussels, Belgium}}, vol.~9899, 2016,
  \href{https://doi.org/10.1117/12.2231095}{DOI}.

\bibitem{Davide}
K.~Abe et~al., \emph{{T2K ND280 Upgrade - Technical Design Report}},
  \href{https://arxiv.org/abs/1901.03750}{{\ttfamily 1901.03750}}.

\bibitem{SHiP}
M.~Anelli et~al., \emph{{A facility to Search for Hidden Particles (SHiP) at
  the CERN SPS}},  \href{https://arxiv.org/abs/1504.04956}{{\ttfamily
  1504.04956}}.

\bibitem{SAMPIC}
E.~Delagnes et~al., \emph{{The SAMPIC Waveform and Time to Digital Converter}},
   in \emph{{2014 IEEE Nuclear Science Symposium (2014 NSS/MIC)}}, (Seattle,
  US), Nov., 2014,
  \href{http://hal.in2p3.fr/in2p3-01082061}{http://hal.in2p3.fr/in2p3-01082061}.

\bibitem{Mineev:2018ekk}
O.~Mineev et~al., \emph{{Beam test results of 3D fine-grained scintillator
  detector prototype for a T2K ND280 neutrino active target}},
  \href{https://arxiv.org/abs/1808.08829}{{\ttfamily 1808.08829}}.

\end{thebibliography}\endgroup
\bibliographystyle{JHEP}
%\begin{thebibliography}{7}
%\bibitem{JINST} C. Betancour et al., Application of large area SiPMs for the readout of a plastic scintillator based timing detector,  JINST 12 (2017) no.11, P11023, arXiv:1709.08972
%\bibitem{Jung} W. Jung et al., Development of TPC Trigger Hodoscope for J-PARC E42/E45 hadron experiment, presented at PD18 in Tokyo, Nov 2018 
%\bibitem{MUSIC} S. Gomez et al., MUSIC: An 8 channel readout ASIC for SiPM arrays, Proc. SPIE 9899, 2016
%\bibitem{Davide} K. Abe et al.,  T2K ND280 Upgrade - Technical Design Report, CERN-SPSC-2019-001, arXiv:1901.03750
%\bibitem{SHiP} M. Anelli et al., A facility to Search for Hidden Particles at the CERN SPS, arXiv:1504.04956, 2015
%\bibitem{SAMPIC} E. Delagnes, D. Breton, H. Grabas, J. Maalmi, P. Rusquart, M. Saimpert, The SAMPIC Waveform and Time to Digital Converter, HAL Id: in2p3-01082061, 2004
%\bibitem{Mineev:2018ekk} O. Mineev et al., Beam test results of 3D fine-grained scintillator detector prototype for a T2K ND280 neutrino active target, arXiv:1808.08829 
%\end{thebibliography}
}

\end{document}